\documentclass[12pt]{article}
\ifx\pdfoutput\undefined
\usepackage[dvips,bookmarks]{hyperref}
\input epsf

\def\figscale#1#2{\epsfxsize=#2\epsfbox{#1.eps}}
\else
\usepackage{hyperref}

\def\figscale#1#2{\pdfximage width#2 {#1.pdf}\pdfrefximage\pdflastximage}
\fi
\usepackage{latexsym}
\hypersetup{colorlinks,bookmarksopen,bookmarksnumbered,citecolor=blue,
    pdfstartview=FitH}
\def\hhref#1{\href{http://arxiv.org/abs/hep-th/#1}{hep-th/#1}}
\def\mhref#1{\href{mailto:#1}{#1}}
\begin{document}
\newcommand{\be}{\begin{equation}}
\newcommand{\ee}{\end{equation}}
\newcommand{\mx}{\mbox}
\newcommand{\mt}{\mathtt}
\newcommand{\p}{\partial}
\newcommand{\st}{\stackrel}
\newcommand{\al}{\alpha}
\newcommand{\bb}{\beta}
\newcommand{\ga}{\gamma}
\newcommand{\te}{\theta}
\newcommand{\de}{\delta}
\newcommand{\et}{\eta}
\newcommand{\ze}{\zeta}
\newcommand{\s}{\sigma}
\newcommand{\e}{\epsilon}
\newcommand{\om}{\omega}
\newcommand{\Om}{\Omega}
\newcommand{\la}{\lambda}
\newcommand{\La}{\Lambda}
\newcommand{\ti}{\widetilde}
\newcommand{\ih}{\hat{i}}
\newcommand{\jh}{\hat{j}}
\newcommand{\kh}{\widehat{k}}
\newcommand{\lh}{\widehat{l}}
\newcommand{\eh}{\widehat{e}}
\newcommand{\ph}{\widehat{p}}
\newcommand{\qh}{\widehat{q}}
\newcommand{\mh}{\widehat{m}}
\newcommand{\nh}{\widehat{n}}
\newcommand{\Dh}{\widehat{D}}
\newcommand{\2}{{\textstyle{1\over 2}}}
\newcommand{\3}{{\textstyle{1\over 3}}}
\newcommand{\4}{{\textstyle{1\over 4}}}
\newcommand{\8}{{\textstyle{1\over 8}}}
\newcommand{\6}{{\textstyle{1\over 16}}}
\newcommand{\ra}{\rightarrow}
\newcommand{\lra}{\longrightarrow}
\newcommand{\Ra}{\Rightarrow}
\newcommand{\im}{\Longleftrightarrow}
\newcommand{\hs}{\hspace{5mm}}
\newcommand{\bea}{\begin{eqnarray}}
\newcommand{\eea}{\end{eqnarray}}
\def\reef#1 #2 {{\it{#1}} {\bf{#2}} }
\thispagestyle{empty} {\bf 1 Sept., 2004 \hfill YITP-SB-04-39}

\vspace{1cm}
\begin{center}{\Large{\bf Linear Regge Trajectories from\\[.2cm]
 Worldsheet Lattice Parton Field
Theory }}\\[1cm]

{\large{\bf Tirthabir Biswas\footnote{\mhref{tirtho@hep.physics.mcgill.ca}},
Marc Grisaru\footnote{\mhref{grisaru@hep.physics.mcgill.ca}}, }}\\[5mm]
{\small Center of High Energy Physics\\
Department of Physics\\
McGill University\\
Montreal, Quebec, Canada}\\[5mm]
{\large{\bf and Warren Siegel\footnote{\mhref{warren@wcgall.physics.sunysb.edu}} }}\\[5mm]
{\small Chen Ning Yang Institute for Theoretical Physics\\
State University of New York\\
Stony Brook, NY, USA}
\end{center}

\begin{abstract}
We show that unlike conventional field theory, the particle field theory of the
string's constituents produces in the ladder approximation linear Regge
trajectories, in accord with its string theory dual. In this theory
propagators are Gaussian and this feature facilitates the perturbative
evaluation of scattering amplitudes. We develop general techniques for studying
their general asymptotic form. We consider radiative corrections
to the ladder Regge trajectory and discover that linearity is lost; however,
this may be due to certain approximations we have made.
\end{abstract}

\newpage
\setcounter{page}{1}

\section{INTRODUCTION}

\noindent In nonrelativistic quantum mechanics the angular momenta and energies
of bound states are related through the language of Regge trajectories
\cite{regge}. The partial wave Schr\"odinger wave function (or the scattering
amplitude) can be continued to complex values of the angular momentum, and for
reasonable potentials has a large domain of meromorphy with poles located at
$J= \alpha (E)$. For values of the energy for which the trajectory function
$\alpha (E)$ passes through nonnegative (half)integers, the value of $J$
corresponds to the angular momentum of a bound state and $E$ is its energy. In
general, for increasing $E$, pole trajectories rise to the right, reach some
maximum, and then return to negative values -- a situation typical of
nonconfining potentials with a finite number of bound states.

In relativistic quantum field theory the scattering amplitude $F(s,t)$  may
also exhibit Regge behavior; with suitable analyticity assumptions the partial
wave amplitudes can be continued to complex values, with pole trajectories $J=
\alpha (s)$ passing through the location of various bound states and
resonances. Furthermore, the Regge poles  also control the high-energy behavior
of the scattering amplitudes in the cross channel, with $F(s,t) \sim  \beta (s)
t^{\alpha (s)}$ as $t \to \infty$ and $s < 0$.  Here the Bethe-Salpeter
equation \cite{bethe} replaces the Schr\"odinger equation as the basic
investigative tool, although its use is relatively limited; the properties of
its solution can only be obtained in certain approximations, such as the ladder
approximation or perturbative Feynman diagram analysis.

 On the experimental side, the data confirms the existence
of families of particles lying on rising trajectories $J= \alpha(s)$ that are
{\em linear}, a fundamental feature, of course, of the Veneziano model and its
stringy cousins. However, as in potential scattering,  in the various
approximations of conventional field theory the trajectories rise for a while
and then fall back towards negative values of $J$ for increasing energy. Thus,
only a few bound states are produced, as characteristic of a Higgs phase;
instead, linearity and an infinite number of bound states are expected to arise
as a consequence of confinement, perhaps due to some infrared catastrophe.
However, such a catastrophe is absent in the usual calculations, which are always
made for massive or off-shell states precisely in order to avoid infrared
divergences.

One approach to nonperturbative string theory is quantization on a suitable
random lattice representing the worldsheet \cite{random}. The lattice is
described by vertices $x_i$ that can be identified with those in a Feynman
diagram of an underlying D-dimensional field theory of ``partons".   The two
theories are ``dual" to each other; what is perturbative in one is
nonperturbative in the other. In particular, since in the string language the
partons are confined, this provides us with a nonperturbative approach to
confinement in the corresponding field theory.

Calculations in the lattice  theory have been  limited mostly to the
two-dimensional gravity aspects of string theory; only the dynamics of the
worldsheet metric has been studied, and relatively little attention has been
devoted to the dynamics of the corresponding partons. In this paper we consider
instead the Feynman diagrams of the particle field theory underlying the bosonic
string.  The mechanism of ``confinement" in this theory differs from that in
ordinary field theories because the propagators are Gaussian \cite{fishnet}
(but see \cite{siegel} for a proposal for a string based on ordinary
propagators).  Both the lack of power-law behavior at large transverse momenta
\cite{veneziano} and the absence of poles in the ``plasma" phase \cite{plasma}
can be understood as unwelcome symptoms of this feature. Nevertheless, a better
understanding of the bound-state mechanism in this model might be helpful in
understanding confinement in quantum chromodynamics, or explaining how the
graviton can arise as a state in a theory whose only fundamental fields are
scalars.

For the bosonic string one starts with the usual functional integral
\be A = \int Dg\ DX\ e^{-S} \ee
of the worldsheet continuum action
\be S = \int{d^2\sigma\over 2\pi}\sqrt{-h}\left[{1\over 2\alpha'}
    h^{\al\bb}(\partial_{\al} X)\cdot (\partial_{\bb} X) +\mu +\kappa R\right] \ee
where the second term is the Liouville (cosmological) term of subcritical
string theory, and the last term is the Euler number, identifying $e^{-\kappa}$
as the string coupling constant.
The worldsheet lattice action is then
\be S' = {1\over 2\alpha'}\sum_{\langle ij\rangle}(x_i -x_j)^2
    +\mu \sum_i 1 +\kappa (Euler) \ee
where ``$\langle ij\rangle$" are the links (edges) of the lattice, ``$i$" are
the vertices, and ``$Euler$" means the Euler number as defined in terms of the
numbers of vertices, edges, and faces.  This action is integrated as
\be A = \sum \int \prod dx\ e^{-S'}= \sum e^{-\mu \sum_i 1}\int  dx\ \prod_{ij}
e^{-\frac{1}{2 \alpha'}  (x_i-x_j)^2 }    \ee
 where the sum over Feynman diagrams replaces
functional integration over the worldsheet metric, and the integration is over
positions of vertices (except for ``external'' vertices which are kept fixed;
 alternatively, the usual external line factors $e^{ikx}$ can be introduced).  Planarity
of the lattice worldsheet is enforced by associating the Feynman diagrams with
a scalar field that is also an $N\times N$ matrix, and using the $1/N$
expansion. This implies the classical identification
\be e^\kappa = N \ee
(which can be modified by worldsheet quantum effects).  The action for this
scalar field is
\be \hat S = \int d^D x\ tr\left[ {1\over 2}\phi e^{-\alpha'\Box/2}\phi
    +G^{n-2}\phi^n \right] \ee
where the kinetic operator has been chosen to agree with the propagator
$exp(-x^2/2\alpha')$ in the amplitude $A$.
The interaction $\phi^n$ has been
chosen arbitrarily; restrictions may follow from consistency of the worldsheet
continuum limit.  Its coefficient is again identified by comparison with the
amplitude:
\be -NG^2 \sim e^{-\mu} \ee
where the proportionality constant depends on normalization of the measure.
(This relation can also be modified by quantum effects.)  Thus the
``perturbative" region of string theory, $\mu\approx 0$, corresponds to the
nonperturbative region of the parton theory, $-NG^2\sim 1$, while conversely the
perturbative region of the parton theory, $NG^2\approx 0$, corresponds to the
nonperturbative region of the string theory, $\mu\approx\infty$.  (There is a
second string-nonperturbative region, $\mu\approx -\infty$, corresponding to
$-NG^2\approx\infty$, which in QCD would be identified with perturbation for the
dual ``monopole" theory.)

In this work we consider 4-point functions in the parton theory with cubic
interaction $\phi^3$. If we choose ``color''
singlets in each of the two pairs of external partons, this corresponds to a
string propagator; equivalently, we are just analyzing a scattering amplitude
to find the Regge trajectory for a two-parton bound state. Another possibility is
to introduce for the external states fundamental-representation ``quarks" in
addition to our adjoint ``gluons", and thus describe open as well as closed
strings.  In the worldsheet continuum limit, taking the ends of the external
propagators on each pair to coincide corresponds to insertion of a ground-state
vertex $exp(ik\cdot X)$.
 (In the early work on Regge theory the interpretation was a bit different in hadronic
      physics, where both the scattered particles and their bound states were hadrons,
       according to the principle of ``nuclear democracy".  Here we go back to the
       original application, where now the constituents are analogous to quarks and
        gluons, and only the bound states are hadrons.)

We will show that ladder graphs are responsible for a Regge
trajectory $\alpha (s)$ that to lowest order in the coupling is linear, as expected
from the string theory
associated with the worldsheet continuum limit. However, including some radiative
corrections seems to spoil the linearity. This may be due to some approximations
we had to make; or it may be that this feature of string theory is not recovered
until one includes more corrections or takes the continuum limit.

\section{REGGE THEORY}

In this section, for the purpose of comparing and contrasting our calculations
with those done in the early days of Regge theory, we summarize some of the
procedures used, mostly in ordinary $\phi^3$ theory, to obtain some information
about pole location and properties. A good review is contained in ref.\ \cite{eden}
and references therein.

Following the original work of Regge, and suggestions that Regge poles might be
relevant for the analysis of high-energy scattering, various results were
obtained on the basis of analyticity assumptions. The main one was the
definition of a suitable continuation to complex angular momentum (the
Froissart-Gribov continuation) of an amplitude which satisfies fixed-energy
dispersion relations \cite{froissart}. It was also shown \cite{gribov} that a
branch cut and/or  essential singularity must be present (at $\ell =-1$ for the
scattering of spinless particles), and that at threshold, $s=4m^2$, an infinite
accumulation of poles occurs at $\ell=-1/2$. Later on more branch cuts were
found from studying particular classes of Feynman diagrams \cite{amati}.

In relativistic field theory the Bethe-Salpeter equation may be used to study
the issue of Regge poles. Lee and Sawyer \cite{lee} considered its continuation
to complex angular momentum, and in ladder approximation (essentially
equivalent to the summation of the diagrams of  Fig.\ 1 in the next section)
established the existence of Regge poles corresponding to bound states. An
equivalent, somewhat simpler, procedure consists in examining directly the
high-energy behavior
 of scattering amplitudes computed in perturbation theory by summing suitable
 sets of Feynman diagrams \cite{federbush}. Since we shall use somewhat similar procedures
 in our present work, we give some details.

One considers an appropriate set of Feynman diagrams (e.g., ladders) for a
two-particle scattering amplitude $A(s,t)$:
\be
A(s,t) = \int d^4k_i \prod_a\frac{1}{{p_a^2 +m^2}} \sim \int  d^4k_i \int_0^{\infty}
\prod_a d\beta_a e^{-\beta_a (p_a^2 +m^2)/2}
\ee
where in our conventions the Mandelstam variables are
$$ s = -(q_1+q_2)^2= -(q_3+q_4)^2,\quad t = -(q_1-q_3)^2;\qquad
    \eta = (-+++) $$
Here $k_i$ are independent loop momenta, and we have introduced Schwinger
parameters to exponentiate the propagators. We note that at this point the only
difference between an ordinary field theory and our  theory is the integration
over the parameters. In our case they are fixed at $\beta_a=\alpha '$.

The Gaussian loop momentum integrations can be carried out, and one is led to
an expression of the form
\be A(s,t) \sim \int_0^{\infty} \prod_a d\beta_a
\frac{N(\beta)}{[C(\beta)]^2} e^{-g(\beta)t -d(s, \beta)}
 \ee
 The large $t$
behavior of the amplitude (which naively vanishes when $t \to \infty$) is
dominated by the neighborhood of points in $\beta$-space where $g(\beta) =0$.
(These points are related to the Landau singularities of the graphs; the
Coleman-Norton interpretation is that they correspond to classical
configurations of point particles, where the Schwinger parameters are their
proper times.)
 Clearly, when  the coefficient of $t$
vanishes one has set to zero parameters which shortcircuit the diagram
(eliminating its $t$ dependence)
 by removing some  of the propagators.
 These can be determined by starting, for example, at the incoming end of the diagram, and
tracing a minimal path (or paths) to the outgoing end,  crossing propagators to
be removed \cite{halliday}. (Compare this with the method we shall use for
similar purposes in Section 4.)  One evaluates the high-energy behavior simply
by setting to zero those $\beta$'s everywhere except in $g(\beta)$ and then
carrying out the integration \cite{federbush}\footnote{Frequently in the
literature the momentum integrals were performed with Feynman parameters, which
correspond to a uniform scaling of all Schwinger parameters. However,
evaluating high-energy behavior requires independent scalings of subsets of the
Schwinger parameters. In these papers, the authors therefore ``unscaled" the
Feynman parameters to re-introduce the Schwinger parameters, then performed the
required scalings.}. Examples and other possibilities are given in \cite{eden},
where diagrams which lead to Regge cuts are also presented.

For the ladder graphs one obtains $g(\beta)=0$  by setting to zero the
parameters which eliminate the rungs. After
setting them to zero everywhere except in $g(\beta)$ the integrations can be
carried out easily and one obtains for the asymptotic behavior of the ladder
with $n$ rungs an expression of the form
 \be
 F_n(s,t) \sim
g^2 \frac{1}{t} [g^2K(s) \ln t]^{n-1}
\ee
where $K(s)$ is just a self-energy
diagram evaluated in two dimensions.  (The power $K^{n-1}$ comes from the fact
that after shortcircuiting the rungs one is left with a product of bubbles.)
Finally, the sum of ladder diagrams gives an asymptotic behavior
\be
\sum
F_n(s,t) = g^2 t^{\alpha (s)},~~~~ \alpha(s) = -1 +g^2K(s)
\ee
For later
comparison we observe that the logarithmic behavior of the individual
contributions and the eventual Regge behavior come from the integration over
Schwinger parameters.

The calculation above gives a Regge trajectory correct to order $g^2$. Higher
order corrections come from considering generalized ladders, which include, in
addition to the single particle exchanges, more complicated ``blobs" inserted
between the simple rungs. The asymptotic behavior is still obtained by
shortcircuiting just the rungs (see ref.\ \cite{eden}, p. 147). Yet another
possibility comes from replacing completely the simple rungs by ``H"-insertions
as in Fig.\ 3(d); two Regge poles are then generated \cite{federbush}.

Although the procedure we have outlined above gives the high $t$ behavior of
each graph directly from the Feynman amplitude, a more useful and often more
powerful method involves the use of the Mellin transform \cite{bjorken}
\be
\ti{F}(z)\equiv \int_0^{\infty}d(-t)\ F(-t)(-t)^{z-1}
\ee
(We use $-t$ as a variable because the Mandelstam variables as usually defined
become negative when continued to Euclidean space, where expressions are most convergent.)
We shall discuss this in
more detail in the next section, but we mention here its main advantage: In
principle it allows the calculation of all the terms in the asymptotic
behavior. As we shall see, it allows easy determination of the asymptotic
behavior of our ladder amplitudes, a task which would be somewhat difficult
otherwise.

Besides the Bethe-Salpeter approach or the investigation of individual Feynman
diagrams, a third method for obtaining some information about Regge
trajectories was provided by the Reggeization program \cite{low}. One
considered the scattering of some elementary particles {\em with spin} (this is
necessary for the program to work) and asked  whether among the
Regge pole bound states one discovered,   the original particles could be
found. (This is connected with the idea of ``bootstrapping" or ``nuclear
democracy" mentioned in the Introduction: There is no distinction between the
elementary particles and their bound states.)  In this program one begins with
partial wave helicity amplitudes computed from tree graphs, and builds up
ladders by using unitarity. For small values of $J$, unphysical helicity
amplitudes (where the helicity exceeds the total angular momentum) have fixed
poles, but through unitarity these fixed poles turn into moving Regge poles,
and the hope was that the original particles lie on the corresponding
trajectories. (Although the use of unitarity is not completely equivalent to
the summation of ladder graphs, calculations seem to indicate that the
difference is irrelevant.)

The Reggeization program was successful by showing that in the case of QED a
Regge pole trajectory in photon-electron scattering does in fact pass, with the
correct quantum numbers and mass, through the position corresponding to the
electron. It failed to show that the photon Reggeizes, and also failed in some
other cases. However, success was achieved with the advent of renormalizable
Yang-Mills theory, when it was shown that the vectors, scalars and fermions
Reggeize \cite{grisaru}. Furthermore, although these theories are not
renormalizable, it was found that in quantum gravity and supergravity certain
necessary conditions for Reggeization of gravitons and gravitini hold
\cite{grisaru'}.

\section{LADDER GRAPHS}

\subsection{Amplitude from Recursion Relations}

We consider an amplitude for
 two incoming particles  with momenta $q_1$ and
$q_2$  and  two   outgoing particles with momenta $q_3$ and $q_4$; the
particles are  off shell. We evaluate the amplitude by solving the
Bethe-Salpeter equation in the ladder approximation through the iteration
procedure depicted in Fig.\ 1. 

\centerline{\figscale{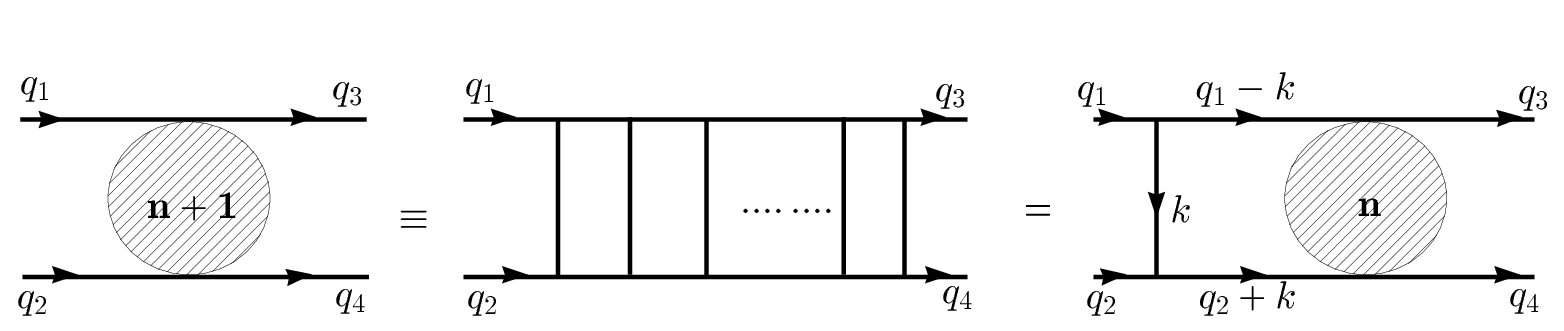}{5.5in}}
\centerline{Fig.\ 1.  Summing ladder diagrams}

\noindent
We denote the $n$-loop diagram by $A_n$. The
propagators are Gaussian, $\exp[ -\2(k_a-k_b)^2]$ and $\exp[ -\2(k_a-q_i)^2]$
(we have set $\alpha' =1$). Upon integration over the $k_i$ one produces, aside
from some numerical factors, exponentials involving squares and scalar products
of the external momenta. Therefore the amplitude will have the form
 \be
A_n=C_n\la^{2(n+1)}\exp[-\2(Q_nq_I^2 + Q_n' q_F^2 -S_ns-T_nt)]\equiv
C_n\la^{2(n+1)}\exp[-\2 E_n] \ee
 where
  \be q_I^2=q_1^2+q_2^2 ~~~,~~ q_F^2 =q_3^2+q_4^2 \quad ; \quad
  q^2 = \sum_{i=1}^4 q_i^2 = q_I^2 + q_F^2
  \ee
and $\la$ is the coupling constant appropriately normalized for the measure $\int d^D x/(2\pi)^{D/2}$ or $\int d^D p/(2\pi)^{D/2}$.  Therefore one can obtain a recursion
relation for the different coefficients by performing the $n+1$st loop
integral. We substitute $A_n$ into the Bethe-Salpeter iteration described by
Fig.\ 1 and obtain
 \be
Q_{n+1}=\frac{Q_n+1}{3+2Q_n+T_n} \ee \be T_{n+1}=\frac{T_n}{3+2Q_n+T_n} \ee
 \be Q'_{n+1}=Q'_n+\frac{T_n(Q_n+1)}{3+2Q_n+T_n} \ee
\be S_{n+1}=S_n+\frac{(Q_n+1)^2}{3+2Q_n+T_n} \ee
 \be
C_{n+1}=\frac{C_n}{(3+2Q_n+T_n)^{D/2}}
\ee
For the ladder diagrams one actually has, by obvious symmetry, $Q'_n=Q_n$ but for
a more general situation with a slightly different input for the right-hand end of the diagram,
they could be different\footnote{For large $n$ the difference between $Q_n$ and $Q'_n$ vanishes
anyway.}. For the time being we have also left the dimension $D$ arbitrary.

To solve the  recursion relations we need initial conditions
which can be obtained  from the tree graph:
$$Q_0=Q'_0=S_0=0$$
\be T_0=C_0=1 \ee
from which it immediately follows that
\be
C_n = T_n^{D/2}
\ee
The linear combination
\be
I_n\equiv3+2Q_n+T_n
\ee
satisfies a simple recursion relation.
\be
I_{n+1}=4-\frac{1}{I_n}
\label{Irec}
\ee
 with the initial condition $I_0 = 4$.
It can be solved to give
 \be
I_n=I_{+}\frac{\tilde{I}x^{2(n+1)}-1}{\tilde{I}x^{2n}-1}
\ee where
$$
I_{\pm}=2\pm\sqrt{3}
$$
are the fixed points of the recursion relation (\ref{Irec}), $\tilde{I}$ is a
constant, and
\be
x\equiv I_-=\frac{1}{I_+}=2-\sqrt{3} \simeq .27 \label{x}
\ee
The recursion relation for $T_n$ can now  be solved as the numerators and the
denominators cancel: \be T_n=\ti{T}\frac{1}{(I_+)^{n}(\tilde{I}x^{2n}-1)}
 \ee
 From the initial conditions we have
 \be
\tilde{I}=x^2 \mbox{ and } \ti{T}=x^2-1
\ee
and the solutions can be rewritten as
\be
I_n=I_+\left(\frac{1-x^{2(n+2)}}{1-x^{2(n+1)}}\right)
 \ee
  and
  \be
T_n=\frac{x^{n}(1-x^2)}{(1-x^{2(n+1)})}
\ee
From these $Q_n$ is determined as well.

 Next, we look at the combination
\be
P_n=4Q_n+4S_n
\ee
The recursion relation for $P_n$ is  simply
 \be
P_{n+1}=P_n+2 \ee once we use the recursion relation for $Q'_{n+1}$ and impose
$Q'_{n+1}=Q_{n+1}$.
 Then
\be
P_n=2n
\ee
Knowing $P_n$ and $Q_n$ we determine
 \be
  S_n=\2\left(n+3+T_n-I_n\right)
 \ee
 which
simplifies to
\be
S_n=\2\left[n+1-\left(\frac{I_+-I_-}{2}\right)\left(\frac{1-x^{n+1}}{1+x^{n+1}}\right)\right]
\ee
Finally, it is easy to obtain
\be
C_n = \left(\frac{x^n(1-x^2)}{1-x^{2(n+1)}}\right)^{\frac{D}{2}}
\ee
where we have used  the initial condition \be C_0=1 \ee

To summarize we have \be A_n=C_n\la^{2(n+1)}e^{-\2 E_n} \ee with

$$E_n=\2q^2n-S_n(s+q^2)-T_nt$$
$$S_n=\2\left[n+1-\sqrt{3}\left(\frac{1-x^{n+1}}{1+x^{n+1}}\right)\right]$$
$$T_n= \frac{x^n(1-x^2)}{(1-x^{2(n+1)})}$$
$$C_n=T_n^{D/2}$$
\be
x=2-\sqrt{3}
\label{xtoyoutoo}
\ee

\subsection{Regge Behavior}

For large, negative $t$ ($t$ is negative by convention for real Euclidean momenta)
the individual amplitudes $A_n$ vanish. However, the asymptotics
 is controlled by their
large $n$ behavior.
Since $x <1$ we have
$$I_n\ra \frac{1}{x}$$
$$T_n\ra (1-x^2)x^{n} $$
$$S_n\ra \2[n+1-\sqrt{3}]$$
$$Q_n\ra \frac{\sqrt{3}-1}{2}$$
\be
C_n\ra [(1-x^2)x^n]^{\frac{D}{2}}
\label{largen}
\ee
Therefore
\be
A_n\simeq \la^{2(n+1)}
x^{\frac{nD}{2}}exp\left[-\2\left(\2(\sqrt{3}-1)
q^2-x^{n}t-\2(n+1-\sqrt{3})s\right)\right]
\ee
It is of the general form (with positive constants   $T$, $\tau$, and $\hat{C}(q^2,s)$,
 given by the above limits)
\be
\hat{A}_n=
\hat{C}\la^{2(n+1)}x^{\frac{nD}{2}}exp\left[\2\left( Tx^{n}t+\tau ns\right)\right]
\ee
The total amplitude has the same asymptotic behavior as the series
\be
\hat{A}=\sum_1^{\infty} \hat{A}_n=\hat{C}\sum_1^{\infty}\la^{2(n+1)}
x^{\frac{nD}{2}}exp\left[\2\left( Tx^{n}t+\tau ns\right)\right]
\label{asbeh}
\ee

To sum the series and exhibit the asymptotic behavior we take, term by term, the Mellin transform
 \be
 \ti{A}(z)\equiv \int_0^{\infty}d(-t)\
\hat{A}(-t)(-t)^{z-1} \ee
with the inverse  transform  given by \be \hat{A}(-t)\equiv
\frac{1}{2\pi i}\int_{b-i\infty}^{b+i\infty}dz\ \ti{A}(z)(-t)^{-z} \ee

We have
\be
\ti{A}_n= \int_0^{\infty}d(-t)\ (-t)^{z-1}\hat{C}\la^{2(n+1)}
x^{\frac{nD}{2}}exp\left[\2\left( Tx^{n-1}t+\tau ns\right)\right]$$
$$=\hat{C}\la^{2(n+1)}x^{\frac{nD}{2}}exp\left[\2\tau ns\right]
\frac{\Gamma(z)}{(\2Tx^{n})^z}
\ee
Summing the series we obtain
$$
\ti{A}=\sum \ti{A}_n=\sum_1^{\infty}
\hat{C}\Gamma(z)\left(\frac{2x}{T}\right)^z\frac{\la^{2(n+1)}
x^{\frac{nD}{2}}e^{\2\tau ns}}{x^{nz}}$$
 \be
=\hat{C'}(s)\Gamma(z)\left(\frac{2x}{T}\right)^z\frac{1}{x^{z}-\la^2
x^{\frac{D}{2}}e^{\2\tau s}} \ee

The asymptotic behavior of the amplitude is
determined by  poles $z_0$ in its Mellin transform \cite{bjorken}:
\be
A(-t)\st{t\ra -\infty}{\longrightarrow}(-t)^{-z_0(s)}\equiv (-t)^{\al(s)}
\ee
where
$\al(s)$ is  the Regge trajectory.

The Mellin transform has a real pole at
$$ z_0=\frac{1}{2\ln x}\tau s+\2 D+2\frac{\ln{\la}}{\ln x}$$
Consequently, we obtain a Regge trajectory
\be
 \al(s)=
\left(\frac{\tau}{-2\ln x}\right) s +\left(-\2 D +2\frac{\ln{\la}}{-\ln x}\right)
\label{traj}
\ee
(For the ladder $\tau = \2$.) Since $x<1$, $\ln{x}$ is negative.
Therefore the trajectory is linear, with positive slope.

The real pole in the Mellin transform gave us the asymptotic
 behavior, but  there are also complex poles located parallel to the imaginary axis at
 \be
  z_n=z_0 + \frac{2\pi in}{\ln x}\equiv Z_R+inZ_I
  \ee
We show in the appendix that  they do not affect the Regge trajectory we have found.

\subsection{Relativistic Harmonic Oscillator}

In this subsection we present an alternative, operatorial, derivation of the
results obtained above. We show that due to the exponential nature of the one-particle
propagators, determining the two-particle propagator  can be
reduced, after a separation of variables, to solving  a harmonic oscillator
problem.

We consider, still in ladder approximation, the two-particle propagator
$\Delta$ (including  the $(2\pi)^D\delta (q_1-q_3) \delta (q_2-q_4)$ term), satisfying
the Bethe-Salpeter equation
\be \Delta = 1 + e^{-H} \Delta \ee
where $e^{-H}$ sticks an extra rung on the sum of ladders  (as in Fig.\ 1).
Explicitly, we can write
\be  e^{-H} = (\hbox{rung propagator})\times(\hbox{two ``side" propagators}) \ee
with integration over either loop momentum (in momentum space) or positions of
vertices (in coordinate space). The propagator is given by
\be \Delta = {1\over 1 -e^{-H}}= \sum(e^{-H})^n   \ee

 The Bethe-Salpeter equation corresponds to
perturbatively solving a Schr\"odinger equation with ``free" Hamiltonian 1 and
potential $-e^{-H}$ and vanishing total energy. Thus, the Schr\"odinger equation
(on the wave function) is
\be 1 - e^{-H} = 0 \quad\to\quad H = 0 \ee
Since $H$ is essentially the sum of $p^2$'s or $x^2$'s for the different
propagators, it is separable (unlike in usual field theory) into
``center-of-mass" and relative pieces. The trivial center-of-mass term
corresponds to the simple
$-ns/2$ dependence in $E_n$.

Explicitly, we want to replace integrals with operator expressions.  In
coordinate space, adding a rung is simply multiplication by the propagator,
while adding the two propagators on the side of the ladder involves integration.
In momentum space, the reverse is true (because duality between vertices and
loops corresponds to Fourier transformation):  Adding the two side propagators is
just multiplication, while adding the rung involves integration.  So, in
operator language adding the rung is simple in terms of the position operators,
while adding the two sides is simple in terms of the momentum operators.  Thus,
adding the two sides followed by adding the rung is performed by the operator
\be e^{-H} = e^{-(x_1-x_2)^2/2}e^{-(p_1^2+p_2^2)/2} \ee
where the $p$'s and $x$'s are now the operators for the two particles.
Separating into average and relative coordinates,
\be p_{1,2} = \2 P \pm p , \quad x_{1,2} = X \pm \2 x \ee
this becomes
\be e^{-H} = e^{-x^2/2}e^{-P^2/4+p^2} \ee
We can now separate out the $P^2=-s$ from the relative parts:
\be e^{-H} = e^{-x^2/2}e^{-p^2}e^{s/4} \ee

It is convenient to use a Hermitian expression by a similarity transformation
that puts half of one exponential on each side, as
\be e^{-H} \quad\to\quad  e^{s/4}e^{-x^2/4}e^{-p^2}e^{-x^2/4}
\quad or \quad e^{s/4}e^{-p^2/2}e^{-x^2/2}e^{-p^2/2} \label{herm}\ee
Since we work in
momentum space, we will use the latter choice, whose similarity transformation
involves only momentum operators. To determine $H$ we need to combine the
 exponentials into a single one.  Since the Baker-Campbell-Haussdorf
theorem requires using only commutators, it is useful to note that the
exponents satisfy the commutation relations of raising and lowering operators,
and we can use the representation
\be \2 x^2 \to \pmatrix{ 0 & 1 \cr 0 & 0 \cr}, \quad
\2 p^2 \to \pmatrix{ 0 & 0 \cr 1 & 0 \cr}, \quad i \2 \{x,p\} \to \pmatrix{ 1 &
0 \cr 0 & -1 \cr} \ee
 So we need in general to evaluate expressions of the form
\be e^{-\alpha p^2/2}e^{-\beta x^2/2}e^{-\alpha p^2/2}\quad\to\quad
e^{-\left({0\atop\alpha}{0\atop 0}\right)}e^{-\left({0\atop 0}{\beta\atop
0}\right)} e^{-\left({0\atop\alpha}{0\atop 0}\right)}   \ee
\be =
\pmatrix{ 1 & 0 \cr -\alpha & 1 \cr}\pmatrix{ 1 & -\beta \cr 0 & 1 \cr}
\pmatrix{ 1 & 0 \cr -\alpha & 1 \cr} = \pmatrix{ 1+\alpha\beta & -\beta \cr
-\alpha(2+\alpha\beta) & 1+\alpha\beta \cr} = e^{-\left({0\atop b}{a\atop
0}\right)} \nonumber \ee
 We determine $a$, $b$ using
\be
\pmatrix{ C & -A \cr -B & C \cr} =
e^{-\left({0\atop b}{a\atop 0}\right)} = cosh(\sqrt{ab}) -
{sinh(\sqrt{ab})\over \sqrt{ab}}\pmatrix{ 0 & a \cr b & 0 \cr} \nonumber \ee
\be \Rightarrow\quad \sqrt{ab} = ln(C+\sqrt{AB}) , \quad
\sqrt{a\over b} = \sqrt{A\over B} \ee
We find from (\ref{herm}), in harmonic oscillator notation,
\be H = -\4 s -ln(\lambda^2) +\omega \left( m \omega \2 x^2 +{1\over m \omega}\2 p^2
\right) \ee
\be \omega = \sqrt{ab} =
ln\left(1+\alpha\beta+\sqrt{(1+\alpha\beta)^2-1}\right) , \quad m \omega =
\sqrt{a\over b} = {\beta\over \sqrt{(1+\alpha\beta)^2-1}} \nonumber \ee
where we have
restored the coupling dependence.  In the present case,
\be \alpha = \beta = 1 \quad\Rightarrow\quad
\omega = ln(2+\sqrt 3) , \quad m\omega = {1 \over\sqrt 3} \ee
By similar manipulations, the first choice of $H$ above (similarity transformation
using $x$'s instead of $p$'s) gives the same result, but with
\be m \omega = {\sqrt 3\over 2} \ee

First we look at just the spectrum, and note that $e^{-H}-1=0$ is the same as $H=2 \pi i n$.
(But the propagators for the two are different, from inverting these
operators.)   We recognize the harmonic oscillator as
a $D$-vector, exactly like the oscillators in the usual string Hamiltonian (but
now we have only one such vector).\footnote{The interpretation of the single
 $D$-vector harmonic oscillator is as follows: The
positions of the two particles in the Bethe-Salpeter equation are two adjacent
points on the random lattice, and the relative coordinate represents the
first derivative of $x(\sigma)$, which corresponds to the first oscillator in
the expansion of $x(\sigma)$.  (A similar model was considered in ref.\
\cite{fineman}.)
  } We can thus identify the ``energy" of the
harmonic oscillator Hamiltonian $m\omega^2\2 x^2 +(1/m)\2 p^2$ as
$(J+D/2)\omega$, where $(D/2)\omega$ is the ground-state energy of the $D$
oscillators, and we identify the integer excitation $J$ with the (maximum) spin
for that energy (from acting with $J$ vector oscillators on the vacuum).  The
result is then
\be 2\pi in = -\4 s -ln(\lambda^2) +ln(2+\sqrt 3)(J+\2 D) \ee
so for the trajectory $J= \alpha (s)$ we have
\be \alpha(s) = -\2 D +{1\over ln(2+\sqrt 3)}[\4 s +ln(\lambda^2) +2\pi in] \ee
in agreement with the result of the previous subsection.

We can determine now the two-particle propagator as
 $ \langle q_3,q_4| \Delta| q_1,q_2 \rangle$ i.e.
\be \langle q_3,q_4| [1-e^{-H}]^{-1}| q_1,q_2 \rangle = \sum_n\langle q_3,q_4|
e^{-nH}|q_1,q_2 \rangle \ee
We use the explicit $H$
\be H = -\4 s -ln(\lambda^2) +ln(2+\sqrt 3)\left({1\over \sqrt 3} \2 x^2
+\sqrt 3\2 p^2 \right) \ee
Because of the similarity transformation, this
corresponds to half (really the square root) of the side-of-the-ladder
propagators on either side of $e^{-H}$.   (If we had used the $x$
transformation, we would instead have half of the rung on either side, which is
harder to fix in momentum space.)  Since we ultimately want ladders with
amputated external propagators, we amputate the other half of the initial and
final propagators on our expression to get for each term in the sum
\be A_{n-1} = \lambda^{2n}e^{(ns+q^2)/4}\langle q_3, q_4 | e^{-nH_0} | q_1, q_2 \rangle \ee
where $A_{n-1}$ is the amplitude for the graph with $n$ rungs ($n-1$ loops) and $H_0$
consists of just the harmonic
oscillator terms ($x^2$ and $p^2$).  (The no-rung graph is a $\delta$-function,
with amputation factors for non-existent propagators.)  The propagator for a
harmonic oscillator in $D$ dimensions (just the product of $D$ one-dimensional ones), with
Wick-rotated time, in momentum space, is
\be \langle q_3, q_4 | e^{-nH_0} | q_1, q_2 \rangle =
[m\omega\ sinh(n\omega)]^{-D/2} e^{-[(p_I^2+p_F^2)cosh(n\omega) -2p_I\cdot
p_F]/2m\omega\ sinh(n\omega)} \ee
where the ``time" is now the integer $n$.
(This differs from the coordinate space one by $x\to p$ and $m\omega\to
1/m\omega$.  There is also the usual momentum conservation $\delta$ function
$(2\pi)^{D/2}\delta^D(q_1+q_2-q_3-q_4)$.)  We now need to use
\be p_I = \2 (q_1-q_2) , \quad p_F = \2 (q_3-q_4) \ee
\be \Rightarrow\quad p_I^2+p_F^2 = \2 (s+q^2) , \quad 2p_I\cdot p_F = \2 (s+2t+q^2) \nonumber \ee
Putting this together (with $\omega=ln(2+\sqrt 3)$ and $m\omega=1/\sqrt 3$)
gives the result of subsection 2.1 (where $x=e^{-\omega}$).

As another interesting example, consider a ``cylindrical" ladder: again two long
lines for the sides of the ladder, but with circular rungs, equivalent to
double rungs, from a $g\phi^4$ coupling instead of $g\phi^3$.  The only
difference in the above calculation is the replacement of the $e^{-x^2}$ factor
with $e^{-2x^2}$.  The only difference in the result is
\be \omega = ln(3+2\sqrt 2) , \quad m\omega = {1\over\sqrt 2} \ee
In particular, this gives a different Regge slope.

\section{GENERAL GRAPHS}

We have shown by exact calculations that  ladder graphs
 corresponding to a Gaussian field theory indeed
produce a linear Regge trajectory, but what can be said about more general graphs?
Do they give rise to radiative corrections to the Regge trajectory we found for the simple ladders
and do they give rise to additional Regge poles?
Is it possible to see that ladder graphs  give the leading
asymptotic behavior for large $t$? To try to answer any of these questions we
have to be able to compute the asymptotic behavior of an amplitude coming from an arbitrary
graph. As compared to
old calculations on Regge behavior we are helped by two features: a) The diagrams
in our theory are all planar and b) the dependence on $s$ and $t$ is always exponential
for any graph. We  show  in this section how this dependence can be determined in principle
 in terms of the adjacency
matrix of the dual graph. The more difficult task, how to determine  the
 asymptotic behavior of sums of general graphs, is beyond the scope of this paper.

\subsection{Adjacency Matrix, Edge and Path Weights}

We start by looking at the dual  graph of a given momentum space graph  where each
simplex is replaced   by a point corresponding to a    loop momentum, $k_a$.\footnote{In
the literature one often chooses to start with triangulation of the worldsheet, then dualizes
to $\phi^3$
Feynman diagrams.  We choose to start instead  with the Feynman diagrams
because it relates directly to our calculation, has the more physical interpretation,
and the unitary choice of integration measure is obvious from the usual Feynman rules.}
In the case of the ladder this is shown in Fig.\ 2.

\vskip-.5in
$$ \figscale{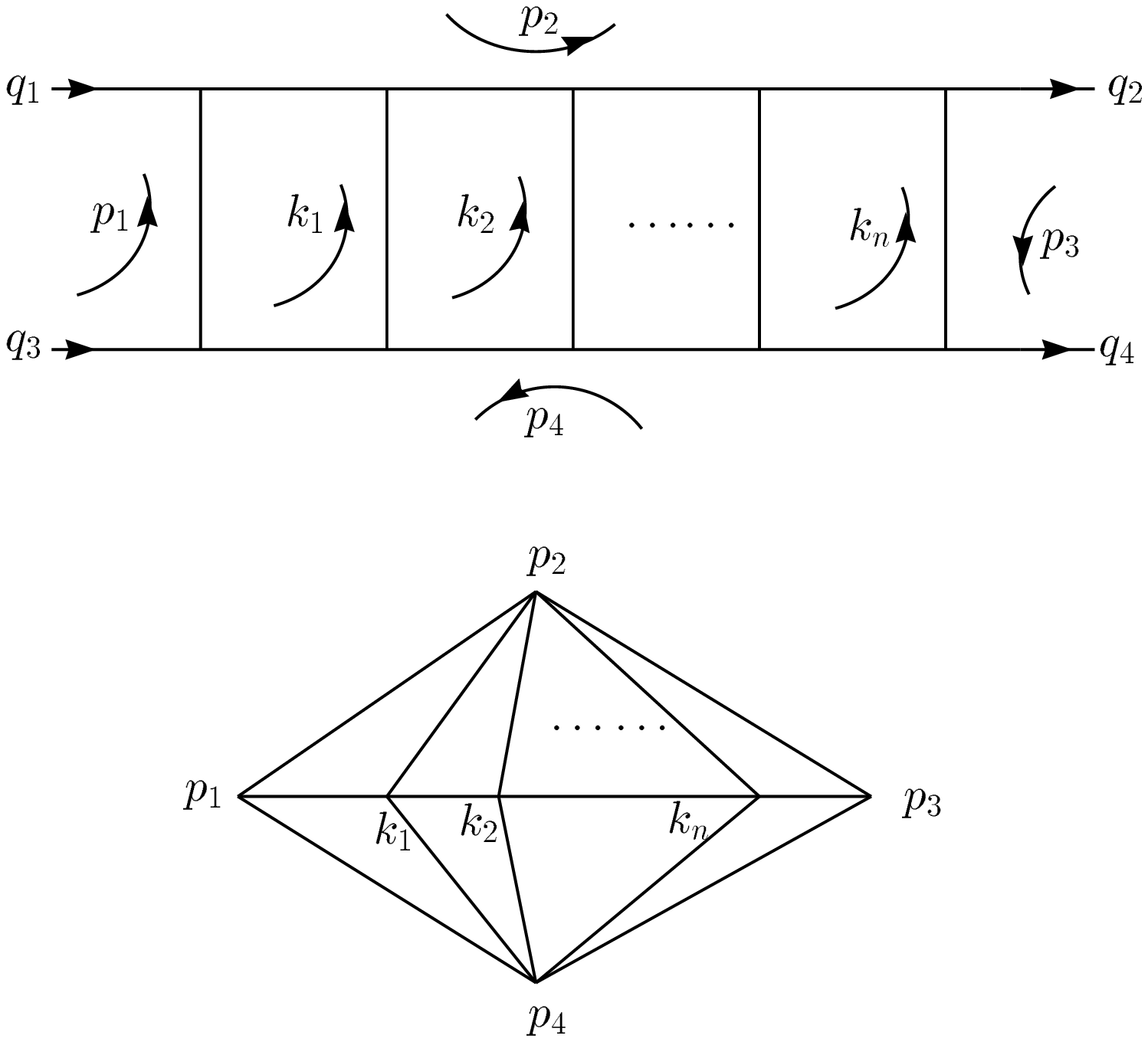}{4.0in} $$
\centerline{Fig.\ 2.  Momentum labeling and dual diagram}

The external four points correspond to the external loop momenta,
which we label as $p_i, \ i=1,...,4$. They are related to the  external
particle momenta in the following way
$$q_1=p_2-p_1$$
$$q_2=p_1-p_4$$
$$q_3=p_2-p_3$$
\be q_4=p_3-p_4
\label{loop}\ee
The Mandelstam variables are
\bea
&&s \equiv -(q_1+q_2)^2 = -(p_2-p_4)^2 \nonumber\\
&& t \equiv -(q_3-q_1)^2 = -(p_1-p_3)^2
\eea
We note the ``gauge'' invariance $ p_i \ra p_i +r$.

Now, if in the original graph  two simplices (loops) are adjacent to each other then
in the dual graph they are connected by an edge.
Thus the dual graph is bounded by four edges
connecting the $p_i$ along with  internal momentum points $k_{a}$
  variously connected to each other. Let $A^{\ih\jh} = (0,1,2, ...)$  denote the adjacency matrix
  in the dual graph, where $\ih=i,a$.
 Our first objective is to compute the exponent $E$ corresponding to a particular graph.
 (For the time being we will not use the
   Einstein summation convention unless stated explicitly.) We have
\be
E=\frac{1}{2}\sum_{\ih\jh}A^{\ih\jh}(p_{\ih}-p_{\jh})^2=\2\sum_{ij}A^{ij}(p_i-p_j)^2
+\2\sum_{ab}A^{ab}(k_{a}-k_{b})^2+\sum_{ib}A^{ib}(p_{i}-k_{b})^2
\ee
The first term is a constant, $E_1$, the same for all  graphs. Expanding
the second and  third terms we find
\bea
E & = & E_1+   \sum_{ab}A^{ab}k_b^2-\sum_{ab}A^{ab}k_a \cdot k_b +\sum_{ib}A^{ib}p_{i}^2+
\sum_{ib}A^{ib}k_b^2-2\sum_{ib}A^{ib}p_i \cdot k_b \nonumber \\
& = & E_1 +\sum_{i}p_{i}^2\left(\sum_{b}A^{ib}\right)+  \sum_{b}k_b^2\left(\sum_{\ih}A^{\ih b}\right)-
\left[\sum_{ab}A^{ab}k_a \cdot k_b -2\sum_{ib}A^{ib}p_i \cdot k_b \right] \nonumber \\
&&
\eea
The second term in the last equation contains only squares of the external loop momenta and we will
soon argue that for large $t$ behavior it is unimportant. For the third
term we note that the sum within parentheses is the ``degree"
$d_{b}$ of the internal momentum (the number of lines meeting at the point $k_b$ or,
in the original diagram, the number of propagator lines bounding the corresponding loop;
 $d=4$ for the ladder diagram). We have
\be
 E = E_1 +E_2(p_i^2)+  \sum_{b}k_b^2d_{b}-\left[\sum_{ab}A^{ab}k_a \cdot k_b-2\sum_{ib}A^{ib}p_i
 \cdot k_b\right]\ee

Let us  rescale the internal momenta and the adjacency matrix as
 \be
  k_a\ra
\sqrt{d_a}k_a \ee \be A^{ab}\ra \frac{A^{ab}}{\sqrt{d_ad_b}}~~~, ~~~~
A^{ib}\ra \frac{A^{ib}}{\sqrt{d_b}}
 \ee
The exponent
becomes\footnote{Since we are only interested in the dependence of the external
momenta we do not keep track of the Jacobian.}
\be
E=E_1 +E_2(p_i^2)+  \sum_{b}(\de^{ab}-
A^{ab})k_a \cdot k_b -2\sum_{b}k_b \cdot \left(\sum_{i}A^{ib}p_{i}\right)
 \ee
This procedure essentially attaches a ``weight''  to each edge.
For the case of the ladder diagram the rescaled adjacency matrix has the following appearance
 \be A=\left( \begin{array}{ccccc}
0 & a& 0 &0 &..\\
a & 0& a &0 &..\\
0 & a& 0 &a &..\\
0 & 0& a &0 &..\\
..&..&..&.. &..
\end{array} \right)
\label{adjacency}
\ee and $a=1/4$ is the edge weight.

Our next task is to perform the (Gaussian) loop integrals. Using (we reintroduce
the summation convention)
\be
\int [dx] e^{-\2(A^{ij}x^ix^j+B_ix^i)}\sim e^{-\2\ti{E}}~~~,~~~~
\ti{E}=-\4B_lA^{lk}B_k \ee
we obtain, after some simplifications,
 \be \ti{E}=E_1
+E_2(p_i^2)-p_iA^{ic}(I- A)^{-1}_{cd}A^{d j}p_j\equiv E_1 +E_2(p_i^2)+E_3(p_i)
\label{pathweight}
\ee

We can now expand the third term in a power series of the rescaled adjacency
matrix
\be E_3=-p_iA^{ic}(\de^{cd}+ A^{cd}+A^{cc'}A^{c'd}+\dots)A^{d j}\cdot p_j
\ee
and its interpretation is  clear. It picks up a contribution only when there
exists a path between the external points $p_i$ and $p_j$, with the contribution
becoming smaller and smaller with each additional internal point introduced in
the path since the edge weights are less than 1. Henceforth we refer to the product
of all the edge weights along a path as the ``path weight".
The coefficient of
$p_ip_j\ (i\neq j)$ in the exponent $E_3$  is then given
by the sum of all path weights connecting $p_i$ and $p_j$ (including smaller and smaller
contributions from paths that
involve back and forth retracing through the vertices).

Let us choose ({\em cf.}\ eq.\ (\ref{loop})) the ``gauge'' $p_2=0$. In this gauge
\bea
&&p_1 \cdot p_3=\2(q_1^2+q_3^2+t) \nonumber\\
&&p_1 \cdot p_4=\2(q_1^2-s-q_2^2)\nonumber \\
&&p_3 \cdot p_4=\2(q_3^2-s-q_4^2)
\eea
We note that, with the possible exception of
the tree graphs, $E_1$ and $E_2$ do not contain any $p_i \cdot p_j$ cross term.
Therefore, in this gauge the coefficient of $t$ in the exponent of the amplitude is
directly proportional to the
coefficient of $p_1 \cdot p_3$ and we have, showing just these terms,
\be
\tilde E = -A^{1c}(\delta^{cd}  +A^{cd}+...)A^{d3}p_1 \cdot p_3 +...
    = -\2 A^{1c}(\delta^{cd}  +A^{cd}+A^{cf}A^{fd}+ ...)A^{d3} t + ...
\label{totalpath}\ee i.e., the coefficient is $-\2$ the total path weight
between the points $p_1$ and $p_3$. Obviously this is a gauge independent
statement. In a similar fashion, picking instead the gauge $p_4=0$, one can
show that the coefficient of $s$ is $-\2$ the total path weight between the
points $p_2$ and $p_4$.

Thus we have a simple expression for the $t$ and $s$
exponents in terms of the rescaled adjacency matrix or edge and path weights, providing us
with significant insight into the amplitude contributions coming from arbitrary
graphs. For example, using the notion of path weights it is  clear
why  at each loop level  (i.e., fixed number of internal points) the ladder
graphs have the leading high  $t$ behavior, i.e., smallest [$p_1$,$p_3$] path
contribution. There is only a single path connecting $p_1$ to
$p_3$ and moreover it contains all the points in the path making the path
weight the smallest\footnote{There are some complications coming from
subleading paths which allows for back and forth motion. Also, some of the
edge-weights in a general graph can be smaller than that of the ladder graphs
because some of the points can have $d>4$. However, one can still argue that the
net path-weight for the ladder graphs is indeed the smallest.}.
 Thus it may be
tempting to conclude that the large $t$ asymptotic behavior will indeed be
dominated by the ladder graph contributions. Unfortunately, while summing an
infinite number of graphs the  properties of the amplitude may change,
 potentially  invalidating such an argument. Thus we will take up a
more modest position and  compare, in the next subsection, only a particular
 class of graphs  to the ladder graphs.

\subsection{Thick Ladder Diagrams}

We will look at diagrams that can be obtained from the ladder diagrams by
replacing the rungs with more complicated 4-point subdiagrams, such as those depicted in Fig.\ 3,
in other words by making the vertical lines ``thicker''.

\centerline{\figscale{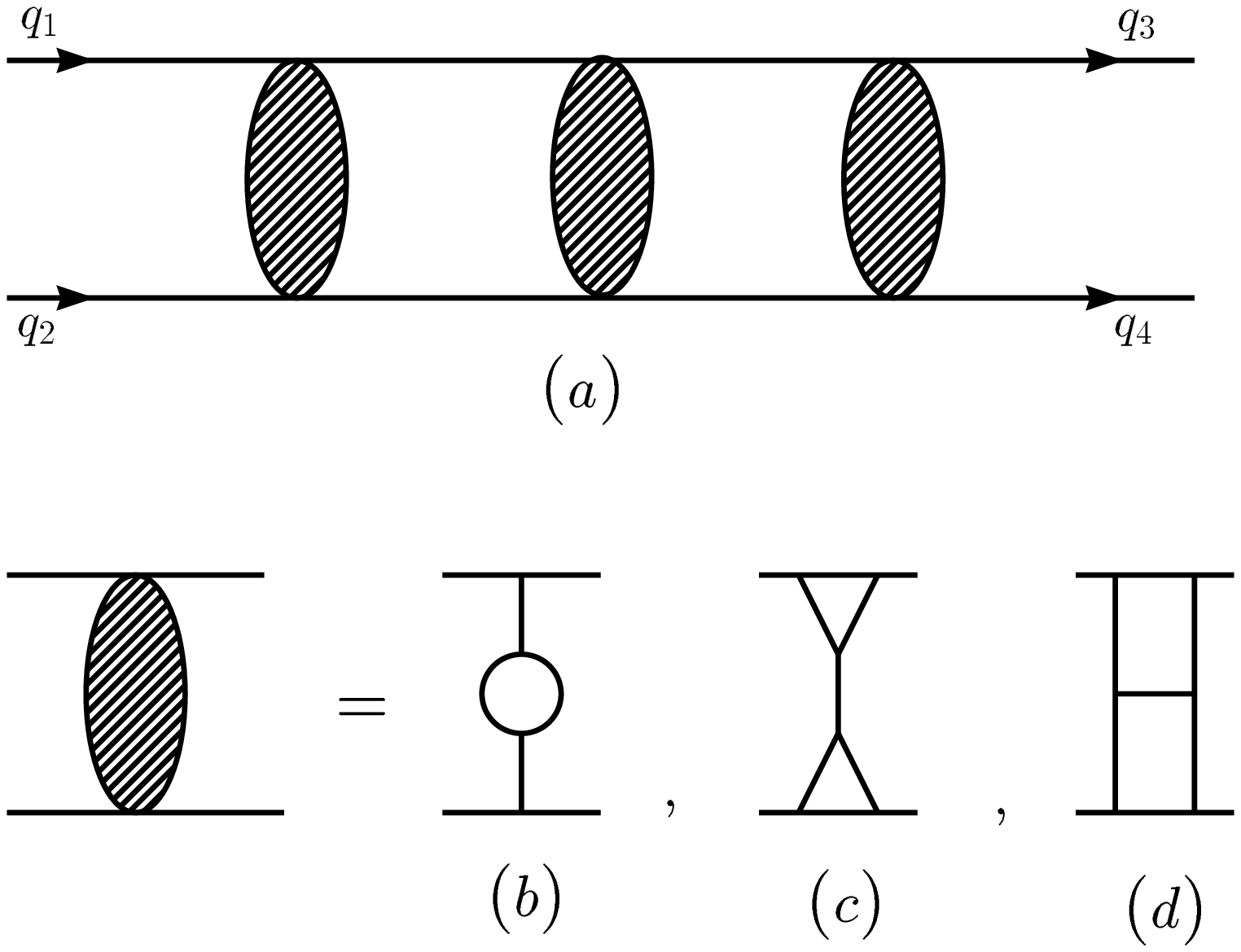}{4.0in}}
\vskip-.7in
\centerline{Fig.\ 3. ``Thick'' ladders}

The method we have used in the previous subsections can be generalized to determine
the asymptotic behavior of these diagrams.  Because of the Gaussian
nature of propagators the exact  expressions for the subdiagrams can be obtained by direct
integration; they
are again  exponentials of form similar to the original one for the simple rungs.
They can be  inserted as kernels into the  Bethe-Salpeter
equations and recursion relations similar to the ones we have already considered can be
obtained and  solved.
 In particular the large $n$ behavior of these thick ladders is the same as before.  Only the
values of the constants $\hat{C},T,\tau$ and $x$ change. Also, the power of
$\la$ associated with the $n$th diagram increases
\be \la^2\ra \la^{2(1+\de)}
\ee
where $\de$ is the number of loops in the subgraph.
One obtains
again Regge behavior of the form
\be
A(-t)\st{t\ra -\infty}{\longrightarrow} (-t)^{\al(s)}
 \ee
 with
 \be
 \al(s)=
\left(\frac{\tau}{-2\ln x}\right) s +\left(-\2 D +2\frac{(1+\de)\ln{\la}}{-\ln x}\right)
\ee
where the values of $x$ and $\tau$
 are given in the table below.

\begin{center}

\begin{tabular}{||l|l|l|l|l||}
\hline
{\bf Case}& {\bf Type}& {\bf x}& {\bf $\tau$} & {\bf slope}\\   \hline

(a)       & Ladder      & .27        & 1/2 & 0.191  \\ \hline

(b)       & Propagator  & .38        & 1/2  & 0.250 \\ \hline

(c)       & Vertex      & .27        & 5/6   & 0.318  \\ \hline

(d)       & {\rm H}-Ladder    & .15        & 9/10   & 0.237 \\ \hline
\end{tabular}

\end{center}

We also have given the numerical values of the Regge slopes from each
 class of ladders. However, on their own they don't have
much significance;  one should consider instead diagrams which
are combinations with thin and thick rungs,  so as to provide
 radiative corrections to the pure ladder trajectories. (We consider this in
 the next section.) We have presented them here for comparison with the
  approximate results we can obtain
from the adjacency matrix methods that we consider now.

As a warm-up exercise let us again look at
a ladder graph. There is only one path connecting $p_1$ and $p_3$ and it includes
  all the points in the path. For a ladder with  $n$-loops, or $n$ edges
with edge-weight $a \equiv a_L=1/4$ (the loops are bounded by four propagators;
equivalently, in the dual diagram, four edges meet at one point) the path-weight $P$
is  naively given by
\be
 P=a_{\mt{L}}^{n}=(.25)^{n}
 \label{shortest}
 \ee
    One immediately notices  a discrepancy between (\ref{shortest})
and the exact result (\ref{x}) directly obtained using recursion relations. This
is because we have considered only the ``shortest-path'' contribution
($A_{p_1k_1}A_{k_1k_2} A_{k_2k_3}...A_{k_np_3}$) without including contributions
coming from paths which allow for  back and forth motion along the edges. However, these subleading
 contributions can be computed and  only  rescale the
edge-weight giving the correct value (\ref{largen})
\be
P= x_{\mt{L}}^{n}\simeq (.27)^{n}
 \ee
We will  discuss this rescaling due to back and forth motion  in more detail in
the next section.

Let us  now replace the vertical lines of the ladder graph with something more complicated
 {\em e.g.}, incorporating vertex or propagator corrections such as those in Fig.\
3.  We will refer to the
simplices (loops) separated by the thick lines as big simplices or ``b-simplices''.
For the example with vertex corrections, Fig.\ 3c, the shortest path approximation
gives for $n$ b-simplices
\be
 P=\left(\frac{1}{8}\right)^n =(.125)^{n}\equiv \tilde{a}_{\mt{V}}^{n}
\label{vertex}
\ee
where the factor 8 comes from the fact that the b-simplices (loops separated by the vertical rungs)
have 8 edges. This expression should be compared with the corresponding expression
 for ladders (\ref{shortest}); the edge-weight $a_{\mt{L}}$ has been replaced by an
 effective edge-weight $\tilde{a}_{\mt{V}}$ between the b-simplices.
 (Actually the $a_{\mt{V}}$ obtained from (\ref{vertex}) is only a rather crude approximation
 to the exact
  effective edge weight $a_{\mt{V}}$ between the b-simplices. One should include contributions
  from paths that cut through the top and bottom triangles,
  as well as some back and forth motions to be described later.) We will show in the next
 section that such a replacement with an exact effective edge-weight works for any general thick
 ladder graph,  even when one accounts for all possible paths,  not just the shortest one.

For the  propagator corrections in Fig.\ 3b, we similarly find
\be
\tilde{a}_{\mt{P}}=\frac{2}{8}=0.25
 \ee
 (the 2 because there are two shortest paths) and for the H diagram in Fig.\ 3d
\be
 \tilde{a}_{\mt{H}}=\frac{2}{6.4}\approx 0.08
 \label{H}
 \ee
As we will explain in the next section, the back and forth motion between the b-simplices
 further rescales the effective edge weights $(a\ra x)$ in a manner similar to the ladders.

\section{RADIATIVE CORRECTIONS}

It is evident that one consequence of having exponential propagators is that
for individual graphs the dependence on $s$ and $t$ will always be exponential,
$\sim \exp(Ss +Tt)$. The coefficients $S,T$ are computable exactly in
principle, and approximately in terms of truncations to  short path weights.
For example, in the case of the thick ladders in Fig.\ 3, we could add to the
shortest path contribution computed above also contributions from paths which
enter the small loops in the thick rungs but  do not include paths involving
back and forth motions. Finally, we could include back and forth motions,
within a thick rung or within different b-simplices. In this section we will
define the exact edge weight $a_\mt{E}$ of a thick rung and  show how to
pass from $a_\mt{E}$ to a corresponding $x_\mt{E}$ which gives $T =
x_\mt{E}^n$. We will then use the analysis to estimate the radiative
corrections to the ladder Regge trajectory from
 two classes of thick subdiagrams.

\subsection{Relating Recursion Relations and Path Weights}

Can one explain the transition $a\ra x$ by properly accounting for all the
paths for a general diagram?  We discuss this here, starting again with the ladder graph
with $n$ loops and the matrix
\be
 1- A=\left( \begin{array}{ccccc}
1 & -a& 0 &0 &..\\
-a & 1& -a &0 &..\\
0 & -a& 1 &-a &..\\
0 & 0& -a &1 &..\\
..&..&..&.. &..
\end{array} \right)
\ee
 where $A$ is the rescaled adjacency matrix  of eq.\ (\ref{adjacency}) and
  $a$ is the edge-weight. For ladder graphs $a=1/4$ but as we will see, for thick
 ladders $a$ will be given by the effective edge-weight between two b-simplices.

 Let $\Delta_n$ denote
the determinant of $I-A$. Then, according to (\ref{pathweight}) the exact path weight
(aside from the ``external" factors $A^{1c}$, $A^{d3}$ in (\ref{totalpath}) is
\be P= (I-A)^{-1}_{1n}=\frac{a^{n-1}}{\Delta_n}\ee
The determinant satisfies a Fibonacci type recursion relation
\be
\Delta_n=\Delta_{n-1}-a^2\Delta_{n-2}
\ee
 The most general solution has the form
 \be
\Delta_n=c_+\mu_{+}^n+c_-\mu_-^n
\ee
with
\be
\mu_{\pm}=\frac{1\pm\sqrt{1-4a^2}}{2}
\label{mu}
\ee
Initial conditions determine 
\be
c_\pm=\2\left(1\pm\frac{1}{\sqrt{1-4a^2}}\right)
\ee.

The roots satisfy
\be
\mu_+\mu_-=a^2
\ee
For large $n$ clearly $\mu_+$ dominates $\Delta_n$(recall also that the large $t$ behavior is
dominated by the smallest $P$) so that we have
\be
P\sim\frac{a^n}{\mu_{+}^n}=\left(\frac{\mu_-}{a}\right)^n
\label{genx}
\ee
Specifically for
ladder graphs we have
\be \frac{\mu_-}{a}=\frac{\4(2-\sqrt{3})}{\4}=x \ee
and  we reproduce our previous result (\ref{x}). Thus,
 we have learned how to pass from the shortest path approximation (just crossing edges with edge
 weight $a_\mt{L}$ and a total weight $a_\mt{L}^n$)  to the exact  value $x_\mt{L}$ that
 determines the
$t$-coefficient $T = x^n$.

Let us now investigate how one goes from thin to thick lines. Suppose there are $n$
 b-simplices. Note that any path
from  1 to $n$ can be broken up into steps of one (i.e., a step from one
b-simplex to one of its neighbours). For simplicity we focus first on a path
which does not contain any back and forth motion between  {\em different} b-simplices but only,
possibly, within the
 ``thick'' rungs bordering a given b-simplex.
The path weight  is then
given by
\be
P= \prod_{1}^{n-1}P_{i,i+1}
\ee
where $P_{i,i+1}$ consists of products of edge-weights in the sub-path from $i$ to $i+1$.
These edge weights are formed by starting in the $i$th b-simplex, entering thick rungs either to its
left or right, proceeding back and forth anywhere within these regions and eventually emerging
in  the $i+1$st b-simplex; see Fig.\ 4 for examples.

\vspace{0.1in}

\centerline{\figscale{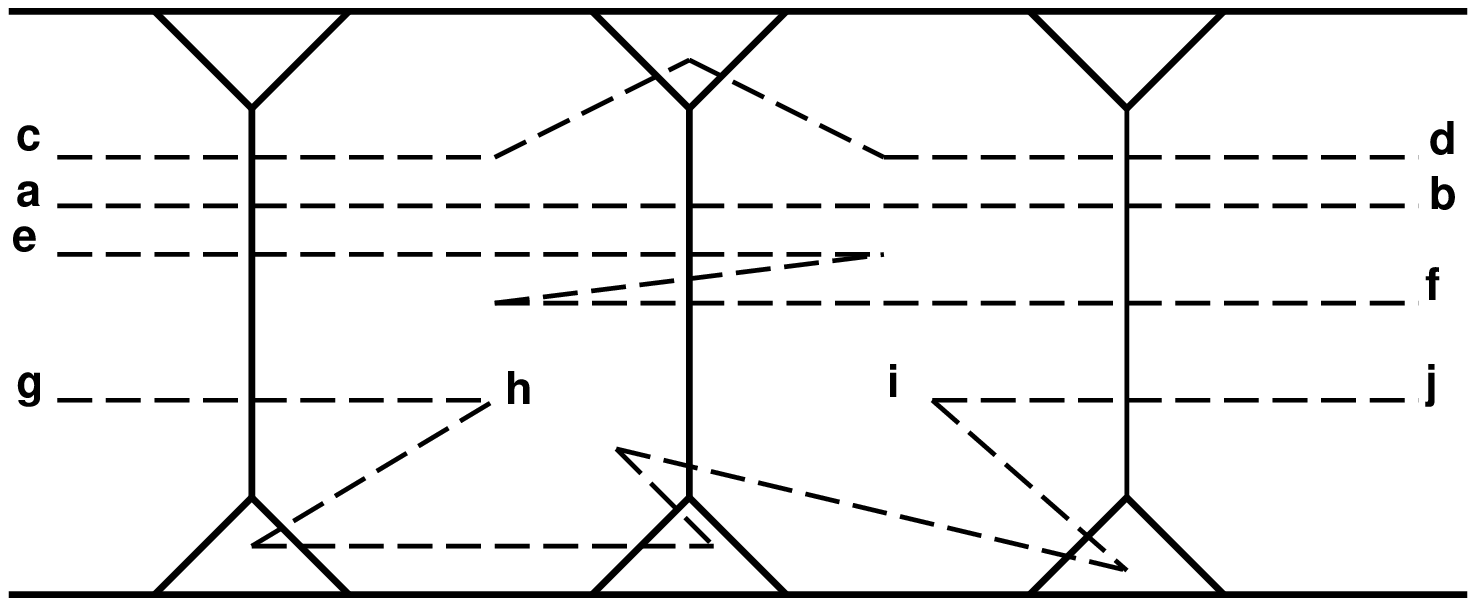}{4.5in}}
 \noindent Fig.\ 4.  Illustrating various paths: (ab) -- shortest path;
 (cd) -- a longer path  ; (ef) -- a path involving back and forth motion;
(gj) -- a more complicated path; the weight of the subpath (hi) is an element
of the effective edge weight of the middle ``thick'' rung.

\vspace{0.1in}

Now, consider a very similar path which only differs from the earlier path in how it goes
from, say, $1$ to $ 2$. The path weight for such a path will be given by
\be
P'= P'_{1,2}\prod_{2}^{n-1}P_{i,i+1}
\ee
so that the sum of the two paths yields
\be
P+P'= (P_{1,2}+P'_{1,2})\prod_{2}^{n-1}P_{i,i+1}
\ee
It is clear that this process can be continued to include all the subpaths originating in 1
and ending in 2, so that
\be
 P+P'+\dots=a_{\mt{e}12}\prod_{2}^{n-1}P_{i,i+1}
 \ee
where $a_{\mt{e}12}$, the sum  of all subpath weights between 1 and 2, is defined to
 be the effective edge-weight between 1 and 2. It is also clear that this process can be
 carried out between any two adjacent b-simplices. Thus the sum of all paths with no back
  and forth motion between the b-simplices is given by
\be
\sum P=\prod_{1}^{n-1}a_{\mt{e}i,i+1}
\label{effective}
\ee

We have defined the effective edge weight $a_{\mt{e}i,i+1}$  as the sum of all
path-weights originating in $i$ and ending in $i+1$ without encountering any other,
i.e., $i-1$ or $i+1$, b-simplex. Essentially in obtaining (\ref{effective}) we have
replaced the sum of products with a product of sums. It is evident that one
can do this for any arbitrary path, i.e., also for those which contains back and forth
motion between the b-simplices themselves. Thus the problem essentially reduces to the
original ladder graph problem except that now we have replaced $a_{\mt{L}}\ra a_{\mt{e}}$.

For the graphs in Fig.\ 3 all the  $a_{\mt{e}}$'s
are the same and the expressions (\ref{vertex}-\ref{H}) give us their shortest path approximations.
 It is clear now that  once the $a_{\mt{e}}$ have been determined (exactly or approximately),
 going over to the corresponding $x$ proceeds in exactly the same
 way as for the
ladder graphs (\ref{mu}-\ref{genx}); in the adjacency matrix one simply replaces the ladder
 edge weights by the effective edge weights. In this fashion we obtain the exact (or approximate,
 if the effective edge weight has been computed approximately)
 asymptotic behavior of any thick ladder graph.

\subsection{Corrected trajectories}

In principle, the Bethe-Salpeter equation provides us with a means of obtaining the exact
scattering amplitude and therefore the exact form of the Regge trajectories.
One only needs to know the kernel function, which consists of the sum of all 2PI
diagrams, planar for our theory. The simplest approximation to the kernel is just
the exchange diagram that leads to pure ladders, and a better approximation would be
obtained by adding to it the simplest radiative corrections corresponding to Figs.\ 3b,c.

\centerline{\figscale{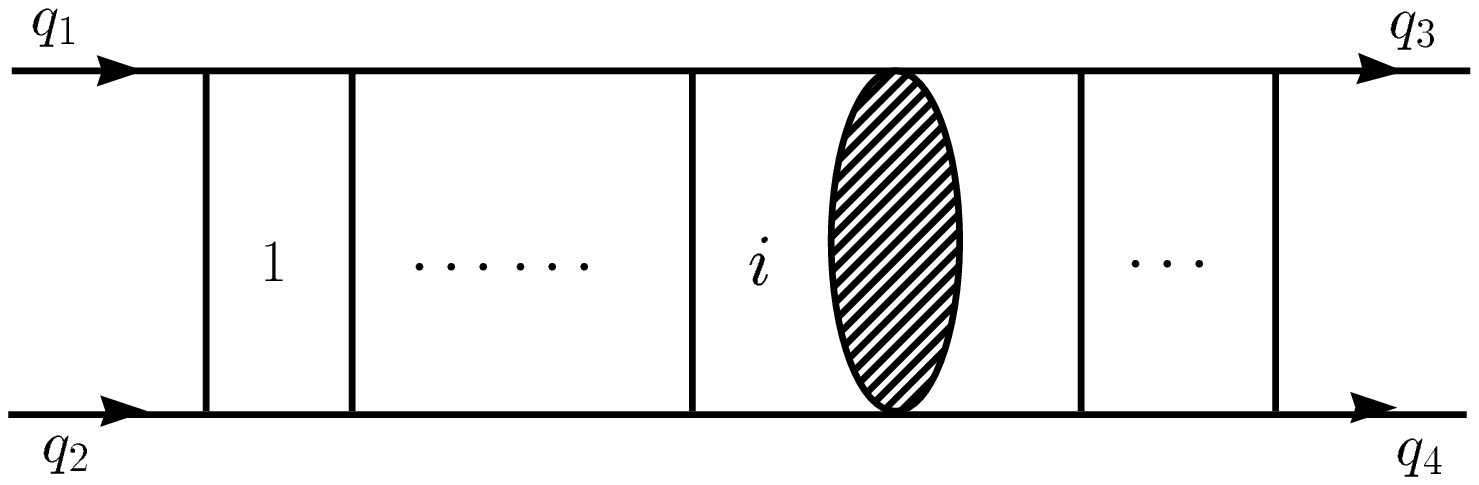}{4in}}
\centerline{Fig.\ 5.  Thick insertion at i-th position}

Let us start with a  ladder graph with $n$ internal loops  or equivalently $n+1$
vertical lines\footnote{There are $n-1$ internal vertical lines, and
the two boundary lines effectively act as an extra edge weight.}.  Now suppose
that we ``thicken'' the $i$-th rung as shown in Fig.\ 5 by replacing it with a
complicated subgraph. The ladder edge weight has
to be then replaced by the effective edge weight for the $i$-th line as
discussed above,
\be a_{\mt{L}}\ra a_{\mt{e}}\ee
so that in the ``shortest'' path approximation the path weight becomes
\be P_{n}= a_{\mt{L}}^{n-1}a_{\mt{e}}\ee
We expect this to get modified once the back and forth motions are
included
\be P_{n}\ra P_{n}= x^{n-1}y \ee
where $x$ and $y$ are the rescaled edge weights for the ladder and the
thick line respectively. (This is not exact; see the discussion below.)
In a similar fashion the path weight which determines the coefficient of
the Mandelstam variable $s$ will change. Since in eq.\ (\ref{asbeh}) the factor $n$
in the coefficient counts the number of simple paths from $p_2$ to $p_4$ we might guess
that $\tau n \longrightarrow \tau (n-1) +\s$. (Again, we don't expect this to be the full story.)
Thus we may guess that the corresponding amplitude looks like
\be\sim \la^{2(n+\de)}e^{\2\{x^{n-1}yt+[\tau (n-1)+\s]s \} }\ee
There are $n$ places where such a  subgraph could be inserted. So the
overall contribution to the amplitude is
\be A_{n,1}\sim n\la^{2(n+\de)}e^{\2(x^{n-1}yt+\tau ns)}\ee

We can extend the argument to $p$ insertions:
\be
 P_{n,p}\simeq x^{n-p}y^p
 \label{approx}
 \ee
One can have these $p$ insertions in $(^n_p)$ ways so that
\be
A_{n,p}=\left(^n_p\right)\la^{2(n+p\de)}e^{\2\{ Tx^{n-p}y^pt+[\tau(n-p)+\s
p]s\}  }
\ee
$$= \la^{2n} e^{\2\tau s} \left(^n_p\right)\la^{2p\de}
e^{\2 Tx^{n-p}y^pt}e^{\2 (\sigma -\tau )ps} $$
Then taking the Mellin
transformation we get \be \ti{A}_{n,p}=\la^{2n} e^{\2\tau n s} \left(^n_p
\right) \frac{\Gamma(z)}{\left(\2 Tx^{n-p}y^p\right)^z} \la^{2p\de}
\left(e^{\2(\sigma -\tau ) s}\right)^p \ee Summing over $p$  gives \be
\ti{A}_{n}=  \Gamma(z)\left( \frac{ 2 }{  T  }\right)^z \frac{\la^{2n}
e^{\2\tau ns}}{x^{nz}}\left[1+\left(\frac{x}{y}\right)^z\la^{2\de} e^{\2(\sigma
-\tau ) s}\right]^n \ee One can now also sum over $n$ to get \be \ti{A}=
\Gamma(z)\left( \frac{ 2 }{  T  }\right)^z \left\{ 1- \frac{\la^{2} e^{\2\tau
s}}{x^{z}}\left[1+  \left(\frac{x}{y}\right)^z\la^{2\de} e^{\2(\sigma -\tau )
s}\right]\right\}^{-1} \label{summation}
\ee
The  Mellin transform has a pole (or perhaps more) at
\be 1=\frac{\la^{2} e^{\2\tau s}}{x^{z}}\left[1+
\left(\frac{x}{y}\right)^z\la^{2\de}  e^{\2(\sigma -\tau ) s}\right]
\ee
or
\bea z \ln x &=& \ln \lambda^2 + \frac{1}{2} \tau s + \ln \left[ 1+ \left(
\frac{x}{y}\right)^z
\lambda^{2\delta} e^{\frac{1}{2}(\sigma - \tau )s} \right] \nonumber \\
& \simeq &\ln \lambda^2 + \frac{1}{2} \tau s + \left(\frac{x}{y}\right)^z \lambda^{2 \delta}
e^{\frac{1}{2}(\sigma -\tau )s} \nonumber\\
& \simeq & \ln \lambda^2 + \frac{1}{2} \tau s + \frac{\lambda^{2\delta}}{y^z}
e^{z \ln x +\frac{1}{2}
(\sigma -\tau )s}\nonumber\\
& \simeq & \ln \lambda^2 +\frac{1}{2}\tau s + \lambda^{2(1+\delta )} e^{
\frac{\sigma}{2}s - z \ln y} \eea
where we have expanded the logarithm (for
small coupling constant) and iterated using the small coupling solution.

The radiative corrections seem to spoil the linearity of the original Regge trajectory. However,
 in reaching the result  we have made some approximations
which are not completely justified. For example, we have assumed that the expression in
(\ref{approx}) is valid irrespective of the ordering of thin and thick rungs and this is definitely
not the case since the effective edge weight of a thick line depends on the nature of its
neighbors.

 As a simple exercise we have looked at the exact expression for the path weight in the case of
 $n+1$ ordinary rungs followed by $m$ thick rungs.
The procedure follows that of Subsection 4.1  where now the matrix $1-A$ has
entries with $n$ rows and columns containing the  simple ladder weight $a$  and
$m$ rows and columns where an effective edge weight $d$ for thick ladders
appears, and in between an additional, transitional row and column
 where  different weights yet, $b$ and $c$ appear (because at the loop which is bordered by
 a thin line on one side and a thick line on the other side the effective edge weights
  are different; also, we have ignored a similar effect where the thick rungs end).
 The total path weight, $[(1-A)^{-1}]_{if}$  between the initial and the final rungs can
 be calculated exactly and one finds a result
\be
P\sim x^n y^m \left[1- ba \frac{ \Delta_{n-1}(a)}{\Delta_n(a)} - cd \frac{
\Delta_{m-1}(d)}{\Delta_m(d)}\right]^{-1}
\label{deviation}
\ee
 where, as before,
$x=a^n/\Delta_n(a)$ and $y= d^m/\Delta_m(d)$ and the $\Delta$'s are the
corresponding determinants for the $n \times n$ and $m \times m$ submatrices.

This is the simplest case and it already shows deviations from the approximate
expression we have assumed in (\ref{approx})\footnote{Note, for large values of $n,m$ the raios  $\Delta_{n-1}(a)/\Delta_n(a)$ and $\Delta_{m-1}(d)/\Delta_m(d)$ are constants and then indeed the expression reduces to (\ref{approx}), but they depend on $n,m$ when their  values are small.}; if the two kinds of rungs are
intermingled the result is expected to be different as well. Furthermore, the
expression we assumed for the path weight controlling the coefficient of $s$ is
also not accurate. (In the limit of large $n$, $m$ the expression in brackets in
(\ref{deviation}) reduces to a constant so that the overall powers of $x$, $y$ are not affected;
however, similar and independent constants from other orderings of thin and thick ladders could
change the result of the summations that led to (\ref{summation}).)
Thus, it is somewhat difficult without
 further work to judge the reliability of the trajectory corrections we have found.

\section{CONCLUSIONS}

In this work we have studied a model field theory defined by the lattice approximation to
the relativistic string.
Unlike ordinary field theory, one is dealing with exponential propagators and this feature
significantly modifies the usual properties of theories with conventional propagators.
Ultraviolet divergences are in general absent and the calculation of amplitudes is very much
simpler.

We have concentrated on the Regge behavior of scattering amplitudes. By solving the
Bethe-Salpeter equation we showed that in ladder
approximation the four-particle amplitude exhibits Regge behavior with a trajectory which is
a linear function of the energy (see (\ref{traj})). We also presented an alternative, operatorial
method for obtaining this result. We have developed general techniques
for obtaining the
high energy behavior of any diagram and we have attempted to determine the effect
of radiative corrections on the linear trajectory we found for the simple ladder. We found
that these corrections spoil the linearity of the trajectory;
however we used a rather crude approximation and it is conceivable that more precise
calculations would change this conclusion.

There  are other  calculations that could be carried out in the field theory
described in this paper.
The exponential nature of the propagators implies that no UV or IR divergences are encountered
and because of the planar nature of the diagrams  other problems are avoided.
For example, it is not difficult to compute in standard fashion the one-loop effective potential;
 other features can  be investigated as well.

\medskip

\noindent {\large\bf  Acknowledgments}

\smallskip
The work of T. Biswas and M. Grisaru is supported by NSERC Grant No.\ 204540.
The work of M. Grisaru is also supported by NSF Grant No.\ PHY-0070475.
The work of W. Siegel is supported in part by NSF Grant No\ PHY-0354776.

\section{Appendix: Imaginary Poles}

 We have pointed out in the main text the existence of complex poles in the
 Mellin transform, located at $z_n = z_0 + i \frac{2\pi}{\ln x} =Z_R +in Z_I$.
In the inverse Mellin transform
 \be
A(-t)=\frac{1}{2i\pi}\int_{c-i\infty}^{c+i\infty}d(-t)\ (-t)^{-z}\tilde{A}(z)
\ee
each of
them  contributes to the integral when one translates the contour towards the
left in order to obtain the asymptotic
behavior \cite{eden}. We show that these additional pole
contributions do not alter the behavior we found earlier.

We begin by
computing the residues at these poles,
\be
\mt{Res}(A_{z=z_n})=\hat{C'}(s)\Gamma(z_n)\left(\frac{2x}{T}\right)^{z_n}\lim_{z\ra
z_n}\frac{z-z_n}{x^{z}-\la^2 x^{\frac{D}{2}}e^{\2\tau s}} \ee
We evaluate
$$
\lim_{z\ra z_n}\frac{x^{z}-\la^2 x^{\frac{D}{2}}e^{\2\tau
s}}{z-z_n}
=\lim_{z'\ra 0}\frac{x^{z'}x^{z_n}-\la^2 x^{\frac{D}{2}}e^{\2\tau s}}{z'}$$
\be =\la^2 x^{\frac{D}{2}}e^{\2\tau s}\lim_{z'\ra 0}\frac{x^{z'}-1}{z'}
=\la^2 x^{\frac{D}{2}}e^{\2\tau s}\ln x \ee
and obtain
\be \mt{Res}(A_{z=z_n})=\hat{C'}(s)\Gamma(z_n)\left(\frac{2x}{T}\right)^{z_n}
\frac{1}{\la^2 x^{\frac{D}{2}}e^{\2\tau s}\ln x} \ee
Thus  the asymptotic behavior of $A(-t)$ is given by
\be A(-t)\sim \sum_{n=-\infty}^{\infty}(-t)^{-Z_R-inZ_I}\hat{C'}(s)\Gamma(z_n)
\left(\frac{2x}{T}\right)^{z_n}\frac{1}{\la^2 x^{\frac{D}{2}}
e^{\2\tau s}\ln x} = \ee
$$ \left[\hat{C'}(s)\left(\frac{2x}{T}\right)^{Z_R}\frac{1}{\la^2
x^{\frac{D}{2}}e^{\2\tau s}\ln x}\right](-t)^{-Z_R}\sum_{n=-\infty}^{\infty}
(-t)^{-inZ_I}\left(\frac{2x}{T}\right)^{inZ_I}\Gamma(Z_R+inZ_I)$$
We  will show  that, except for the  $n=0$ term, the sum is indeed bounded so that the asymptotic
behavior is as prescribed by the real pole at $z=z_0$.

 We note that
\be \sum_{n=-\infty}^{\infty}(-t)^{-inZ_I}\left(\frac{2x}{T}\right)^{inZ_I}
\Gamma(Z_R+inZ_I)<\sum_{n=-\infty}^{\infty}\left|t^{-inZ_I}\left(\frac{2x}{T}\right)^{inZ_I} \Gamma(Z_R+inZ_I)\right| $$
$$=\sum_{n=-\infty}^{\infty}|\Gamma(Z_R+inZ_I)|\ee
For large $n$
\be |\Gamma(Z_R+inZ_I)| = \sqrt{2\pi}e^{-\frac{\pi^2}{2} |nZ_I|} |nZ_I|^{Z_R-\2} \ee
Therefore the bounding series converges and the imaginary poles simply modify the coefficient
of the Regge-behaved term $(-t)^{\alpha (s)}$.

\end{document}